\documentclass[journal]{IEEEtran}
\usepackage[colorlinks]{hyperref}
\usepackage{cite}
\usepackage{mathrsfs}
\usepackage{amsthm}
\usepackage{etoolbox} \makeatletter \patchcmd{\@makecaption} {\scshape} {} {} {} \makeatother
\usepackage[cmex10]{amsmath}
\usepackage{float}
\usepackage{mathtools}
 \usepackage{siunitx}
\usepackage{array}
\usepackage{eqparbox}
\usepackage{bm}

\usepackage[table,dvipsnames]{xcolor}
\usepackage{multicol,booktabs,tabularx}

\usepackage{cases}
\usepackage{algorithm}
\usepackage{paralist}
\usepackage{algpseudocode}

\usepackage[utf8]{inputenc}
\usepackage[english]{babel}
\usepackage{graphicx}
\usepackage{subfigure}
\usepackage{epstopdf}
\usepackage{epsfig}
\usepackage{amssymb}
\usepackage{array}
\usepackage{multirow}

\usepackage{caption}
\usepackage[numbers,sort&compress]{natbib} 

\begin{document}
\author{Yuan Gao,~\IEEEmembership{Member,~IEEE}, Yiming Liu, Jun Jiang, Jianbo Du,~\IEEEmembership{Member,~IEEE}, Shunqing Zhang,~\IEEEmembership{Senior Member,~IEEE}, Xiaoli Chu,~\IEEEmembership{Senior Member,~IEEE}, Kai-Kit Wong,~\IEEEmembership{Fellow,~IEEE}
\thanks{This work was supported in part by Shanghai Natural Science Foundation under Grant 22ZR1422200, and in part supported by the 6G Science and Technology Innovation and Future Industry Cultivation Special Project of Shanghai Municipal Science and Technology Commission under Grant 24DP1501001.}
\thanks{Yuan Gao, Yiming Liu and Shunqing Zhang are with the School of Communication and Information Engineering, Shanghai University, China, email: gaoyuansie@shu.edu.cn, 1975231949@shu.edu.cn and shunqing@shu.edu.cn.}
\thanks{Jun Jiang is with School of Advanced Technolog, Xi’an Jiaotong-Liverpool University, Suzhou, China, email: Jun.Jiang25@student.xjtlu.edu.cn.}
\thanks{Jianbo Du is with the School of Communication and Information Engineering, Xi'an University of Posts and Telecommunications, Xi'an, China, e-mail: dujianboo@163.com.}
\thanks{Xiaoli Chu is with the Department of Electronic and Electrical Engineering, the University of Sheffield, UK, e-mail: x.chu@sheffield.ac.uk.}
\thanks{K. K. Wong is with the Department of Electronic and Electrical Engineering, University College London, Torrington Place, WC1E 7JE, United Kingdom and he is also affiliated with the Department of Electronic Engineering, Kyung Hee University, Yongin-si, Korea, e-mail: $\rm kai\text{-}kit.wong@ucl.ac.uk$}

}
\title{JEPA-CFM: A Joint Embedding Predictive Architecture-based Channel Foundation Model for Robust Fluid Antenna Systems }

\maketitle

\begin{abstract}
Fluid antenna systems (FAS) have emerged as a promising technology for sixth-generation (6G) wireless networks. By allowing antenna elements to move freely within a compact region, FAS can exploit rich spatial diversity without additional hardware. However, acquiring real-time channel state information (CSI), extrapolating channel values to unmeasured antenna ports, and determining accurate user positions remain major obstacles. These challenges stem mainly from strong spatial correlations within the limited aperture and the scarcity of observable data. To overcome these limitations, this paper introduces joint embedding predictive architecture (JEPA)-based channel foundation model (CFM) specifically designed for FAS. The model adopts JEPA and, during its self-supervised pre-training phase, learns versatile representations by extracting high-level latent embeddings of masked or unobserved channel segments. Unlike conventional approaches that attempt pixel-by-pixel reconstruction of raw CSI coefficients, JEPA-CFM focuses on predicting abstract structures in a compact feature space. The pre-training objective combines three complementary loss terms: the standard masked autoencoder reconstruction loss, the JEPA latent prediction loss, and a sliced isotropic Gaussian regularization (SIGReg) term. Together, these components prevent representation collapse and significantly enhance robustness under severe spatial correlation and highly sparse observations. After pre-training, the encoder is frozen, and lightweight task-specific heads are attached: a decoder for channel extrapolation and a global average pooling layer followed by a multi-layer perceptron regression head for wireless positioning. Extensive simulations in the realistic DeepMIMO urban scenario demonstrate that JEPA-CFM substantially outperforms the conventional masked autoencoder baseline in channel extrapolation, especially under severe sparsity conditions (5\%–15\% known CSI) and across various signal-to-noise ratio (SNR) levels. For wireless positioning, the model further achieves sub-3 m accuracy using only 25\% known CSI, highlighting strong data efficiency and robustness even with limited observations.
\end{abstract}


\IEEEpeerreviewmaketitle

\section{Introduction}
The drive for greater spectral efficiency, energy efficiency, and reliability in sixth-generation (6G) wireless networks has spurred the development of innovative antenna technologies. Among these, fluid antenna systems (FAS) have emerged as a particularly promising advance \cite{hong2025contemporary,bepari2026fluid,liu2025water}. Unlike conventional fixed-position multiple-input multiple-output (MIMO) or reconfigurable intelligent surface (RIS) systems, FAS allows radiating elements to be repositioned continuously or discretely within a compact region. This capability dynamically captures spatial diversity and mitigates deep fades without additional radio-frequency (RF) chains or hardware overhead \cite{new2024tutorial,wong2020fluid}.

This positional flexibility, however, introduces significant challenges in acquiring and utilizing real-time channel state information (CSI). Critical tasks, such as extrapolating channel values to unmeasured antenna ports, selecting optimal ports for transmission, and determining accurate user positions—become major bottlenecks. The difficulties stem mainly from strong spatial correlations within the limited FAS aperture and the prohibitive overhead of exhaustive pilot-based estimation in fast-changing environments \cite{gao2025ssnet,tang2025accurate,wang2025large,gao2026ai}.

Traditional model-based methods for FAS, which rely on stochastic geometry or deterministic ray-tracing, suffer from high computational complexity and poor scalability as the fluid region expands or environmental dynamics intensify. These approaches also struggle to capture the complex, non-stationary channel statistics induced by antenna movement \cite{SSnet2025gao,zhang2024learning}. Data-driven supervised learning offers greater adaptability yet typically requires task-specific neural networks trained on massive labeled datasets for each individual problem (e.g., separate regressors for channel prediction or classification heads for port selection). The result is fragmented architectures with limited cross-task generalization and heightened vulnerability to distribution shifts in real-world deployments \cite{jiang2025towards}. These shortcomings are especially acute in FAS scenarios, where instantaneous channel conditions must be inferred rapidly from sparse observations. This situation underscores the urgent need for a unified self-supervised representation learning framework that can extract generalizable channel embeddings directly from unlabeled data—one area where channel foundation models (CFMs) have shown initial promise but have not yet fully embraced predictive self-supervision.

Existing CFMs are primarily built on generative paradigms, such as masked autoencoders that reconstruct raw channel coefficients, or discriminative paradigms based on contrastive learning. While effective in certain settings, both approaches exhibit notable limitations for wireless channel modeling. Generative methods focus on pixel-level or coefficient-level reconstruction, rendering them sensitive to noise, high-frequency artifacts, and irrelevant signal details that produce suboptimal representations for downstream inference tasks. Discriminative methods, conversely, depend heavily on carefully designed data augmentations and struggle to model the strong spatial correlations and predictive dynamics inherent in fluid antenna channels. Consequently, they often display limited generalization across diverse propagation environments and demand substantial labeled data when adapted to specific FAS tasks such as channel extrapolation, port selection, and positioning.

In contrast, the joint embedding predictive architecture (JEPA) provides a more powerful alternative by predicting abstract latent embeddings of masked or future channel segments in a high-level representation space, rather than reconstructing raw signals or relying on hand-crafted augmentations \cite{balestriero2025lejepa,lecun2022path}. This latent predictive objective naturally discards unpredictable noise and high-frequency components while explicitly learning higher-order semantic structures, including spatial correlations, multi-port dependencies, environmental dynamics, and spatio-temporal evolution. The resulting representations are more robust, data-efficient, and transferable for inference-oriented tasks \cite{balestriero2025lejepa}. In wireless communications, JEPA has demonstrated encouraging early success in pioneering works such as WirelessJEPA for multi-antenna IQ data \cite{chu2026wirelessjepa}, JEPA-MSAC for multimodal sensing-assisted communications \cite{,zheng2026jepamsac}, and extensions for structured CSI dynamics modeling and cooperative perception. Nevertheless, its adoption remains nascent. Most applications target fixed-antenna or discrete MIMO setups rather than continuous-position FAS; predictor networks often require careful regularization to avoid representation collapse under the extreme spatial correlations of compact FAS apertures; and scalability to real-time port dynamics, online adaptation, and validation under realistic 3GPP channel models or hardware-in-the-loop testbeds is still limited. These gaps motivate the present work.

To bridge these limitations, we propose JEPA-CFM, a novel unified framework that extends the CFM paradigm to FAS by adopting JEPA as the core self-supervised encoder. Our primary contributions are threefold and directly address the challenges outlined above.
\begin{itemize}
\item This paper introduces JEPA as the central self-supervised learning paradigm for CFM. Rather than reconstructing raw channel coefficients at the pixel or complex-value level, the model learns to predict high-level latent embeddings of masked or unobserved channel segments within a compact representation space.
\item The pre-training objective integrates three complementary loss terms: the standard masked autoencoder reconstruction loss, the JEPA latent prediction loss, and a sliced isotropic Gaussian regularization (SIGReg) term. This multi-objective formulation not only enforces faithful reconstruction over masked regions but also promotes stable, non-collapsing latent representations, significantly enhancing generalization under the severe spatial correlations and sparse observations typical of compact FAS apertures.
\item Third, extensive simulations in a realistic urban ray-tracing scenario using DeepMIMO rigorously validate the practical superiority of JEPA-CFM. The framework consistently outperforms the conventional masked autoencoder baseline on both channel extrapolation and wireless positioning tasks. It achieves substantially lower extrapolation error under severe pilot sparsity (5 \%–15 \% known CSI) and across realistic SNR levels (0 dB, 10 dB, and 20 dB), while delivering superior positioning accuracy, reaching sub-3 m error at 25 \% known CSI.
\end{itemize}

The remainder of this paper is structured as follows: Section \ref{sec:related_work} provides a comprehensive review of related literature on FAS technologies, channel foundation models, and self-supervised learning frameworks. Section \ref{sec:system_model} presents the system model of FAS and the downstream tasks formulation. Section \ref{sec:proposed_model} details the proposed JEPA-CFM architecture, including the pretraining objective, encoder design, and mathematical formulation of the predictive latent space. Section \ref{sec:simulation_results} elaborates on the simulation settings and analysis. Finally, Section \ref{sec:conclusions} concludes the paper with discussions on limitations and promising directions for future research, such as multi-user extensions and online continual pretraining.

\section{Related works}
\label{sec:related_work}
FAS have attracted rapidly growing research interest owing to their distinctive ability to reposition radiating elements dynamically within a compact region. This section reviews the current state of the art on the three core FAS tasks examined in this paper: channel extrapolation, port selection, and positioning. For each task, we categorize existing solutions into three groups—model-based schemes, task-specific AI schemes, and multi-task AI schemes—while highlighting their respective strengths and limitations. These limitations, taken together, underscore the need for a unified foundation-model approach.
\subsection{Channel extrapolation for FAS}
Model-based schemes for FAS channel extrapolation typically rely on classical statistical channel models, such as Jake’s model \cite{zhang2024learning}, geometry-based stochastic models, or the Karhunen-Loève expansion \cite{wu2026learned}. These approaches exploit spatial correlation within the compact fluid region by formulating extrapolation as an analytic extension problem, using techniques such as linear or nonlinear interpolation, kernel-based methods, or Nyquist-sampling-based reconstruction. However, they offer limited accuracy in non-stationary or rich-scattering environments, as they struggle to capture complex multi-port dependencies and the high-resolution spatial variations induced by continuous antenna positioning.

Task-specific AI schemes have advanced the field by treating channel extrapolation as an image-reconstruction or supervised regression problem. Representative examples include the ConvNeXt-based CANet with context-adaptive and cross-scale attention mechanisms \cite{jin2025context}, SSNet that employs self-supervised masked modeling for flexible and robust extrapolation \cite{gao2025ssnet}, and diffusion-model-based posterior sampling estimators that effectively capture two-dimensional spatial correlation structures \cite{tang2025accurate}. Although these methods achieve superior accuracy over model-based baselines, they require large amounts of task-specific labeled data, exhibit poor generalization across environments, and incur high retraining costs for new FAS configurations.

Multi-task AI schemes remain relatively scarce in this subdomain. Most existing efforts focus on isolated extrapolation modules with limited integration into joint frameworks that simultaneously address port selection or other downstream tasks. Recent designs empowered by large language models begin to explore unified extrapolation within broader FAS optimization pipelines \cite{wang2025large}. Nevertheless, these approaches still lack a truly universal encoder that supports efficient multi-task adaptation using lightweight heads.
\subsection{Port selection for FAS}
Model-based port selection schemes typically optimize criteria such as instantaneous signal-to-noise ratio, channel capacity, or outage probability through exhaustive search, greedy algorithms, or convex relaxations (e.g., joint convex relaxation for fluid-MIMO). These methods provide theoretical insights and optimality guarantees under perfect channel state information but become computationally prohibitive as the number of ports grows and suffer performance degradation under imperfect or partial observations \cite{chai2022port,efrem2024transmit}.

Task-specific AI schemes dominate the recent literature and have made substantial practical advances in FAS port selection. Early foundational work \cite{chai2022port} introduced machine-learning-based fast port selection algorithms that achieve significant outage probability reduction by using only 10\% of the observed ports and exploiting strong spatial correlations. Subsequent efforts include LSTM-based temporal-spatial predictors for efficient channel estimation and port selection, as well as various deep reinforcement learning frameworks for dynamic and switching-cost-aware port selection \cite{liu2026switching}.

More recent developments incorporate graph neural networks for joint port position and beamforming optimization \cite{xu2025toward}, along with large-language-model-assisted predictors such as Port-LLM \cite{zhang2025port} for mobility-aware port selection. These techniques consistently reduce the number of required observed ports to 10-20\% while approaching near-optimal performance in terms of signal-to-noise ratio or sum-rate. However, they are typically designed for single-objective optimization (e.g., signal-to-noise ratio maximization or energy efficiency) and lack inherent cross-task transferability to channel extrapolation or positioning. Although multi-task AI schemes have begun to emerge that jointly optimize port selection with beamforming or precoding in multi-user and integrated sensing and communication scenarios \cite{salem2025secure,zhang2024joint}, they generally rely on separate models or tightly coupled end-to-end networks. Such designs scale poorly and require full retraining when tasks or environments change. Consequently, the absence of a shared, generalizable representation continues to hinder efficient simultaneous handling of port selection alongside channel extrapolation and positioning.
\subsection{Positioning for FAS}
Model-based positioning schemes leverage the spatial diversity of FAS through enhanced angle-of-arrival, time-of-arrival, or received-signal-strength-indicator-based algorithms, often combined with Fisher information analysis to derive fundamental Cramér-Rao lower bounds \cite{salem2025fundamental}. These approaches benefit from the angular diversity gained by fluid port switching but remain sensitive to calibration errors, non-line-of-sight conditions, and the discrete nature of port sampling.

Task-specific AI schemes employ deep learning models, such as deep reinforcement learning for antenna positioning in integrated sensing and communication or unmanned aerial vehicle scenarios \cite{yang2025towards}, received-signal-strength-indicator-based neural networks that exploit inter-port correlation for multi-point maximum-likelihood estimation \cite{liu2025rssi}, and transformer-based collaborative learning for three-dimensional localization. These methods exploit the rich spatial sampling provided by FAS to achieve sub-meter accuracy improvements over fixed-antenna systems. Nevertheless, they are trained end-to-end for localization alone and suffer from high data requirements and limited robustness to distribution shifts.

Multi-task AI schemes in this area are still nascent. A few frameworks integrate positioning with communication tasks (e.g., FAS-assisted integrated sensing and communication), but most still use decoupled modules rather than a unified representation. This fragmentation prevents seamless joint optimization of localization, channel extrapolation, and port selection within a single lightweight architecture \cite{zhang2025indoor}.
\subsection{Limitations of existing research}
Overall, while substantial progress has been made on individual FAS tasks, the literature reveals a clear gap: the predominance of task-specific or narrowly multi-task models that lack a universal, self-supervised encoder capable of supporting multiple downstream tasks with minimal adaptation overhead. This motivates the proposed JEPA-CFM framework.

\section{System model and problem formulation}
\label{sec:system_model}
\subsection{System model}

Consider a single-cell downlink wireless communication scenario wherein a base station (BS) equipped with a conventional uniform linear array (ULA) provides spatial multiplexing services to a mobile user equipment (UE). The BS ULA consists of $M$ fixed-position antennas, where $M = 8$, separated by a normalized inter-element spacing of $d_{\mathrm{BS}} = 0.5\lambda$, with $\lambda$ denoting the operating carrier wavelength. Conversely, the UE is equipped with an innovative two-dimensional FAS, which enables a fluidic radio-frequency (RF) radiating element to dynamically migrate within a continuous rectangular two-dimensional area of physical dimensions $W_{\mathrm{X}} \times W_{\mathrm{Y}}$.

To facilitate digital processing and tractability, the continuous physical aperture of the 2D FAS is discretized into a uniform grid comprising $N_{\mathrm{S}} = N_{\mathrm{X}} \times N_{\mathrm{Y}}$ switchable ports along the horizontal and vertical axes, respectively. In accordance with the concrete implementation parameters, the grid configuration is specified as $N_{\mathrm{X}} = 64$ and $N_{\mathrm{Y}} = 64$, yielding a total of $N_{\mathrm{S}} = 4096$ discrete port locations. The uniform spatial port spacings along the $x$-axis and $y$-axis are explicitly given by
\begin{equation}\label{eq:port_spacing}
d_{\mathrm{X}} = \frac{W_{\mathrm{X}}}{N_{\mathrm{X}} - 1}, \quad d_{\mathrm{Y}} = \frac{W_{\mathrm{Y}}}{N_{\mathrm{Y}} - 1}.
\end{equation}

Let $\bm{p}(i,j) = [ (i-1)d_{\mathrm{X}}, (j-1)d_{\mathrm{Y}}, 0 ]^{\mathrm{T}} \in \mathbb{R}^{3 \times 1}$ represent the three-dimensional spatial coordinates of a port located at the grid index $(i,j)$, where $1 \le i \le N_{\mathrm{X}}$ and $1 \le j \le N_{\mathrm{Y}}$. For the $n$-th channel observation sample, the narrowband complex-valued channel matrix between the BS ULA and the full grid of the UE 2D FAS is denoted as $\bm{H}_n \in \mathbb{C}^{N_{\mathrm{S}} \times M}$. To faithfully encapsulate the site-specific physical propagation characteristics, electromagnetic scattering, and spatial topology, $\bm{H}_n$ is generated via a deterministic ray-tracing simulation framework based on the DeepMIMO O128 dataset. Specifically, the wireless channel is characterized by a multi-path geometric propagation model comprising $L = 5$ distinct physical paths. Let $\beta_{n,\ell} \in \mathbb{C}$ denote the complex path gain of the $\ell$-th multi-path component for the $n$-th sample. The azimuth and elevation angles of arrival (AoA) at the UE for the $\ell$-th path are designated as $\theta_{n,\ell}$ and $\phi_{n,\ell}$, respectively, while the corresponding angle of departure (AoD) from the BS ULA is denoted as $\psi_{n,\ell}$.

The spatial array response vector at the BS ULA for the $\ell$-th path is expressed as
\begin{equation}\label{eq:bs_steering}
\begin{split}
\bm{a}_{\mathrm{BS}}(\psi_{n,\ell}) = \Big[ 1, \exp\left(\mathrm{j} 2\pi d_{\mathrm{BS}} \sin\psi_{n,\ell}\right), \dots, \\
\exp\left(\mathrm{j} 2\pi (M-1) d_{\mathrm{BS}} \sin\psi_{n,\ell}\right) \Big]^{\mathrm{T}},
\end{split}
\end{equation}
where $\mathrm{j} = \sqrt{-1}$ represents the imaginary unit. Concurrently, the spatial steering vector of the 2D FAS at the UE corresponding to the port index $s$, which maps bijectively to the grid coordinates $(i,j)$ via $s = (j-1)N_{\mathrm{X}} + i$, is defined by its components as

\begin{align}
&[\bm{a}_{\mathrm{UE}}(\theta_{n,\ell}, \phi_{n,\ell})]_s =\nonumber\\& \exp \Bigg( \mathrm{j} \frac{2\pi}{\lambda} \bigg( \splitfrac{\sin\theta_{n,\ell}\cos\phi_{n,\ell} (i-1)d_{\mathrm{X}}}{+ \sin\theta_{n,\ell}\sin\phi_{n,\ell} (j-1)d_{\mathrm{Y}}} \bigg) \Bigg).\label{eq:ue_steering}
\end{align}

By accumulating the contributions from all $L$ dominant scattering paths, the individual complex channel coefficient $h_{n}(i, j, m)$ between the $m$-th antenna of the BS ULA and the $(i,j)$-th port of the UE FAS for the $n$-th sample is mathematically formulated as

\begin{align}
&h_{n}(i, j, m) = \nonumber\\&\sum_{\ell=1}^{L} \beta_{n,\ell} \exp \Bigg( \mathrm{j} \frac{2\pi}{\lambda} \bigg( \splitfrac{\sin\theta_{n,\ell}\cos\phi_{n,\ell} (i-1)d_{\mathrm{X}}}{+ \sin\theta_{n,\ell}\sin\phi_{n,\ell} (j-1)d_{\mathrm{Y}}} \bigg) \Bigg) \nonumber\\&
\times \exp \left( -\mathrm{j} 2\pi (m-1) d_{\mathrm{BS}} \sin\psi_{n,\ell} \right),\label{eq:channel_coefficient}
\end{align}
where $1 \le m \le M$. The collective full-grid channel state information (CSI) matrix $\bm{H}_n$ thus establishes a rigid bijective mapping between the discrete physical port arrangements and the spatial-domain multi-path multi-input multi-output (MIMO) channel characteristics.

\subsection{Problem formulation}
AI/ML is considered as a key enabler for 6G, 3GPP has prioritized AI/ML in air interface, including CSI acquisition, wireless positioning and beam management \cite{38_843,gao20263gpp}. 
\subsubsection{CSI acquisition}
Channel extrapolation has shown its great potential to acquire CSI with limited overhead \cite{gao2025enabling}. Channel extrapolation aims to predict unknown CSI in resource elements, subbands, bandwidth parts (BWPs), component carriers (CCs), or future time instances based on limited observed measurements. Channel extrapolation is generally classified into time-domain, frequency and antenna-domain \cite{gao2026ai}. Time-domain CSI extrapolation is particularly critical in mobile communications, especially in scenarios with rapidly changing channels (e.g., high-mobility users), where accurate CSI extrapolation can optimize resource allocation, beamforming, and precoding design. Frequency-domain CSI extrapolation is particularly valuable in scenarios like FDD and multi-band systems, where reducing feedback overhead and enhancing spectral efficiency are critical. Antenna-domain channel extrapolation aims to infer the CSI of unmeasured antennas on the same panel using CSI from a subset of antennas, thereby reducing DL training and feedback overhead.

The objective of channel extrapolation is to find $\mathbb{F}_\mathrm{*}$ to minimize the mean squared error (MSE) between the extrapolated CSI $\widehat{\mathbf{H}}'=\mathbb{F}_\mathrm{*}(\mathbf{H}_\mathrm{known})$ and the ground truth CSI $\mathbf{H}$ in time-domain, frequency-domain, antenna-domain or a mix of above domains, which is formulated as follows:
\begin{equation}
     \min_{\mathbb{F}_\mathrm{*}}(|\widehat{\mathbf{H}'}-\mathbf{H}|^2).
\end{equation}

\subsubsection{Wireless positioning}
In 5G NR networks, positioning reference signals (PRS) are transmitted by one or multiple gNBs with configurable density, periodicity, and bandwidth to enable high-precision downlink positioning through measurements such as Time of Arrival (TOA), Time Difference of Arrival (TDOA), and Angle of Arrival (AOA) \cite{gao2026sidelink}. This positioning task can be naturally formulated as a regression problem, where an AI model directly predicts the UE coordinates from processed PRS observations. Specifically, a neural network \(g_{\theta}\) learns to map input features \(\mathbf{H}_{\text{PRS}}\) (e.g., correlation peaks, channel estimates, or multi-TRP measurement vectors derived from PRS) to the 2D/3D position vector \(\mathbf{p}\). The corresponding optimization problem is given by
\begin{equation}
\min_{\theta} \frac{1}{N} \sum_{i=1}^{N} \left\| g_{\theta}(\mathbf{H}_{\text{PRS}}) - \mathbf{p}_i^* \right\|_2^2, 
\end{equation}where \(\mathbf{p}_i^*\) denotes the ground-truth position, thereby allowing data-driven models to achieve superior accuracy and robustness in multipath and NLOS environments compared to conventional geometric methods.

\section{Proposed model}
\label{sec:proposed_model}
\begin{figure*}[!t]
\centering\includegraphics[width=1\textwidth]{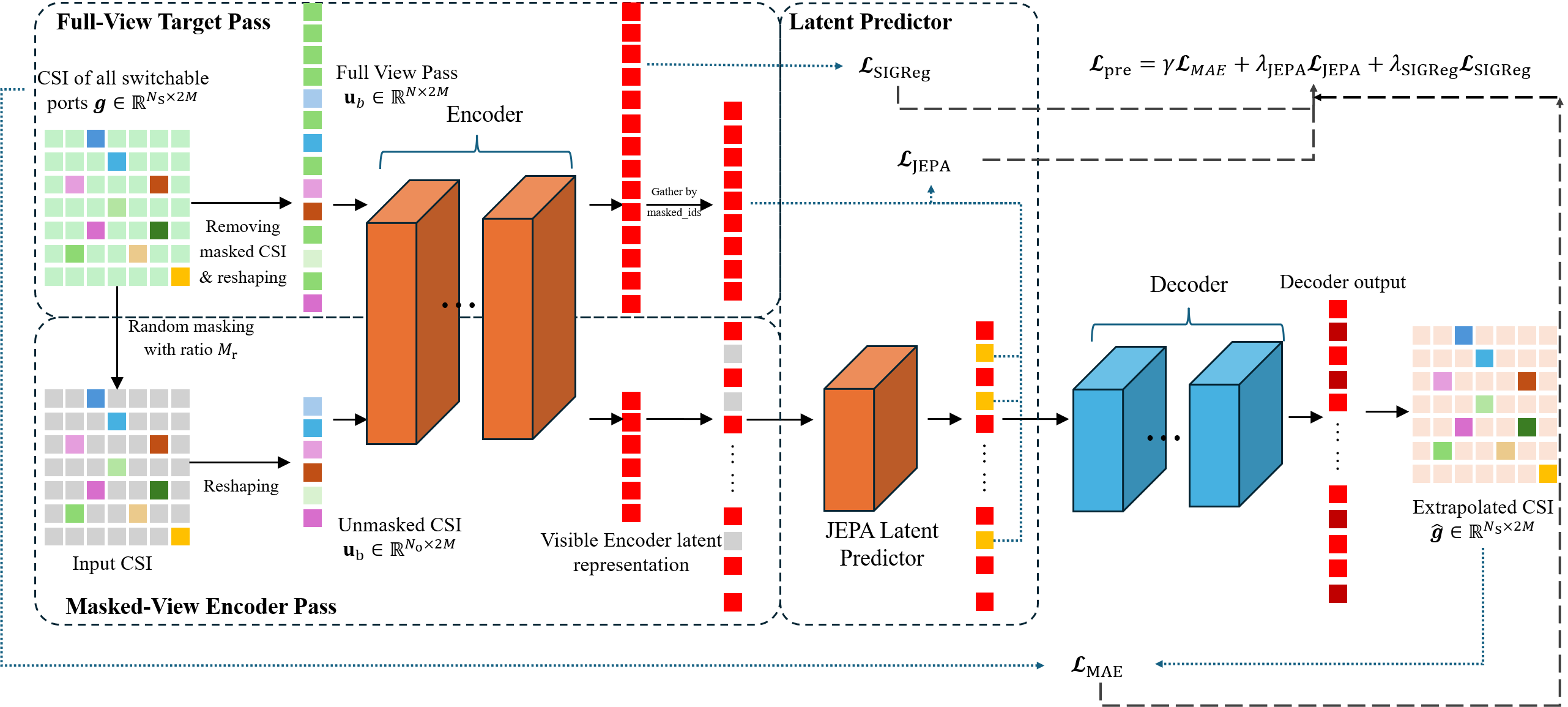}
\caption{The architecture of the proposed model.}
\label{fig:architecture}
\end{figure*}

\begin{figure}[!t]
\centering\includegraphics[width=0.5\textwidth]{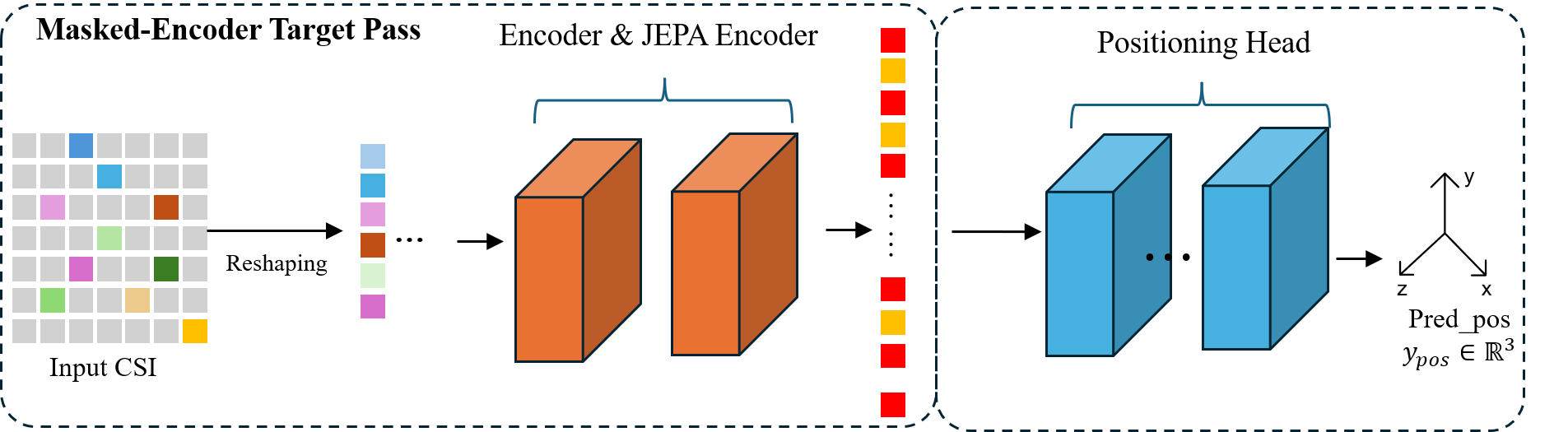}
\caption{The pipeline of using pre-trained encoder and JEPA encoder for downstream task positioning.}
\label{fig:positioning}
\end{figure}

The proposed model is first pre-trained via masked reconstruction as illustrated in Fig. \ref{fig:architecture}, where the novel JEAP loss and sliced isotropic Gaussian regularization (SIGReg) loss are in-cooperated with the reconstruction loss. Specifically, the JEAP loss is proposed to avoid exclusively relying on high-dimensional raw signal reconstruction and obtain general-purpose features for downstream tasks. The SIGReg loss is integrated to avoid representation collapse. We then adopted the pre-training encoder and JEPA encoder for downstream task fine-tuning, which is illustrated in Fig. \ref{fig:positioning}.
\label{sec:proposed}
\subsection{Model architecture}

\subsubsection{CSI Tensorization and Normalization}
To render the high-dimensional complex-valued channel state information compatible with deep vision backbones such as the Vision Transformer, the raw CSI matrix $\bm{H}_n$ must be mapped onto a real-valued spatial tensor framework. Utilizing the individual complex channel coefficients $h_n(i,j,m)$ derived in \eqref{eq:channel_coefficient}, the complex spatial characteristics are first reorganized into a third-order tensor $\bm{\mathcal{H}}_n \in \mathbb{C}^{M \times N_{\mathrm{X}} \times N_{\mathrm{Y}}}$. To project this representation into the real domain $\mathbb{R}$, the complex tensor is decoupled by segregating its real and imaginary parts along the channel dimension, thereby generating an unnormalized real input tensor $\bm{X}_{n,\mathrm{raw}} \in \mathbb{R}^{2M \times N_{\mathrm{X}} \times N_{\mathrm{Y}}}$. Let $c$ denote the channel index of the tensor, where $1 \le c \le 2M$. The tensor splitting and concatenation operations are formally defined as
\begin{equation}\label{eq:tensor_split}
[\bm{X}_{n,\mathrm{raw}}]_{c, i, j} = 
\begin{cases} 
\mathrm{Re}\left(h_n(i, j, c)\right), & 1 \le c \le M, \\
\mathrm{Im}\left(h_n(i, j, c - M)\right), & M < c \le 2M,
\end{cases}
\end{equation}
where $1 \le i \le N_{\mathrm{X}}$ and $1 \le j \le N_{\mathrm{Y}}$, while $\mathrm{Re}(\cdot)$ and $\mathrm{Im}(\cdot)$ denote the operators extracting the real and imaginary parts of a complex scalar, respectively.

To eliminate macroscopic variations stemming from path loss and large-scale shadowing across distinct communication blocks, a sample-wise normalization strategy is applied. This step ensures that the foundation model concentrates on capturing local spatial correlations rather than scaling differences. For each observation, a scalar channel energy scale factor $\alpha_n \in \mathbb{R}$ is computed using the Frobenius norm of $\bm{X}_{n,\mathrm{raw}}$, which is expressed as
\begin{equation}\label{eq:scale_factor}
\alpha_n = \|\bm{X}_{n,\mathrm{raw}}\|_{\mathrm{F}} = \sqrt{\sum_{c=1}^{2M} \sum_{i=1}^{N_{\mathrm{X}}} \sum_{j=1}^{N_{\mathrm{Y}}} \left([\bm{X}_{n,\mathrm{raw}}]_{c, i, j}\right)^2}.
\end{equation}

Furthermore, to prevent numerical instability or division-by-zero errors when the 2D FAS experiences severe localized deep fading, a protective threshold is integrated into the normalization denominator. The standardized dimensionless input tensor $\bm{X}_n \in \mathbb{R}^{2M \times N_{\mathrm{X}} \times N_{\mathrm{Y}}}$ is rigorously constructed as
\begin{equation}\label{eq:normalization}
\bm{X}_n = \frac{\bm{X}_{n,\mathrm{raw}}}{\alpha_n + \epsilon},
\end{equation}
where $\epsilon = 10^{-8}$ denotes the deep-fading protection regularizer. The resulting standardized tensor $\bm{X}_n$ acts as the pristine structural representation utilized during the self-supervised masked pre-training phase as well as the downstream multi-task finetuning evaluations.

\begin{figure}[!t]
\centering\includegraphics[width=0.5\textwidth]{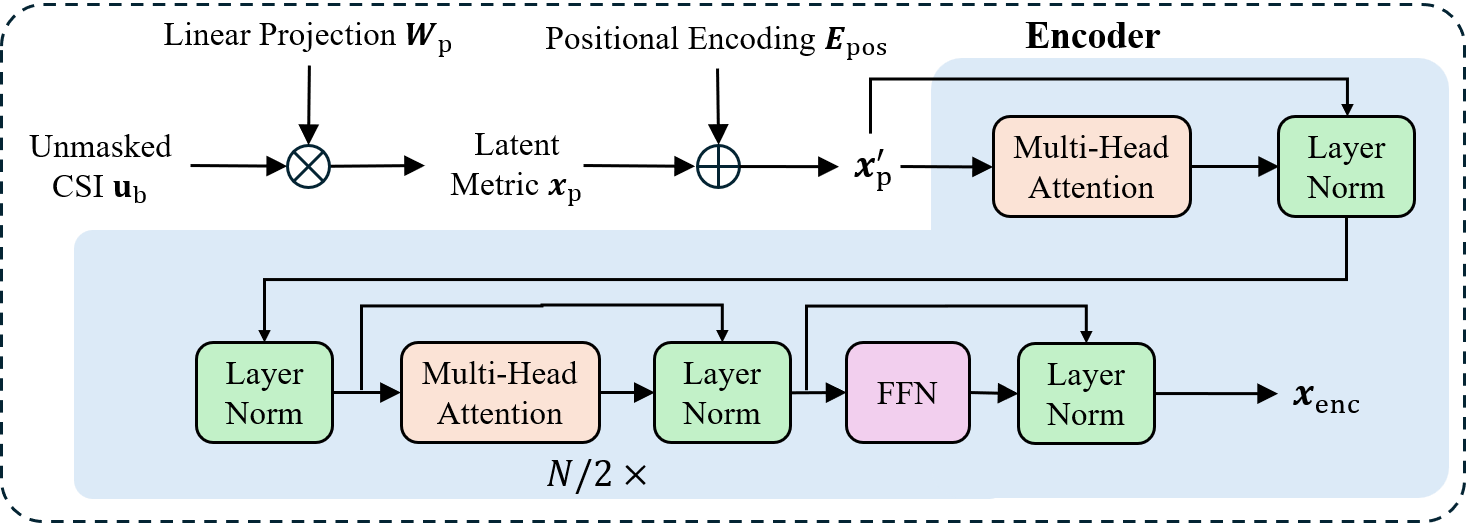}
\caption{The architecture of the encoder.}
\label{fig:encoder}
\end{figure}

\subsubsection{Encoder}
Following the CSI preprocessing and stabilization steps formulated in \eqref{eq:normalization}, the standardized channel state information must be mapped into a high-dimensional latent space to extract robust spatial representations. As illustrated in Fig. \ref{fig:encoder} , to reflect the actual hardware constraints and pilot overhead reductions within the fluid antenna system topology, the encoder operates exclusively on the observed port configurations. Let $\bm{U}_{\mathrm{b}} \in \mathbb{R}^{N_{\mathrm{o}} \times 2M}$ denote the observable port CSI matrix compiled after applying the random spatial mask, where $2$ accounts for the decoupled real and imaginary parts. The parameter $N_{\mathrm{o}}$ defines the number of observable ports remaining under a strict mask ratio $M_{\mathrm{r}}$, such that $N_{\mathrm{o}} = \lfloor N_{\mathrm{S}}(1 - M_{\mathrm{r}}) \rfloor $.

To map these localized channel signatures into the hidden embedding space, the input matrix $\bm{U}_{\mathrm{b}}$ is passed through a learnable linear projection layer characterized by the projection weight matrix $\bm{W}_{\mathrm{p}} \in \mathbb{R}^{2M \times d_{\mathrm{model}}}$, where $d_{\mathrm{model}} = 192$ specifies the encoder hidden dimension. This feature transformation yields the initial latent patch embedding matrix $\bm{X}_{\mathrm{p}} \in \mathbb{R}^{N_{\mathrm{o}} \times d_{\mathrm{model}}}$, which is expressed as
\begin{equation}\label{eq:patch_projection}
\bm{X}_{\mathrm{p}} = \bm{U}_{\mathrm{b}} \bm{W}_{\mathrm{p}}.
\end{equation}

To systematically embed the underlying spatial coordinate relationships of the physical 2D FAS ports into the feature sequence, a channel-model-invariant two-dimensional sine-cosine positional embedding matrix $\bm{E}_{\mathrm{pos}} \in \mathbb{R}^{N_{\mathrm{o}} \times d_{\mathrm{model}}}$ is superposed element-wise. This operation produces the spatially augmented feature matrix $\bm{X}_{\mathrm{p}}' \in \mathbb{R}^{N_{\mathrm{o}} \times d_{\mathrm{model}}}$, which is given by
\begin{equation}\label{eq:add_pos_encoder}
\bm{X}_{\mathrm{p}}' = \bm{X}_{\mathrm{p}} + \bm{E}_{\mathrm{pos}}.
\end{equation}

For any given antenna port situated at grid coordinates $(i,j)$ within the discrete planar surface, its corresponding positional embedding vector $\bm{E}_{\mathrm{pos}}(i,j) \in \mathbb{R}^{1 \times d_{\mathrm{model}}}$ is constructed by concatenating horizontal and vertical frequency components as
\begin{equation}\label{eq:pos_concat}
\bm{E}_{\mathrm{pos}}(i,j) = \mathrm{Concat}\left( \bm{E}_{\mathrm{row}}(i), \bm{E}_{\mathrm{col}}(j) \right),
\end{equation}
where $\bm{E}_{\mathrm{row}}(i) \in \mathbb{R}^{1 \times (d_{\mathrm{model}}/2)}$ and $\bm{E}_{\mathrm{col}}(j) \in \mathbb{R}^{1 \times (d_{\mathrm{model}}/2)}$ represent the horizontal and vertical coordinate encodings, respectively. Let $k$ denote the frequency index variable. The individual elements of the row encoding vector are explicitly defined as
\begin{align}
[\bm{E}_{\mathrm{row}}(i)]_{2k} &= \sin(i \cdot \omega_k), \label{eq:pos_enc_sin} \\
[\bm{E}_{\mathrm{row}}(i)]_{2k+1} &= \cos(i \cdot \omega_k), \label{eq:pos_enc_cos}
\end{align}
where $0 \le k < \lfloor d_{\mathrm{model}}/4 \rfloor$. The angular frequency progression $\omega_k$ is governed by an exponential decay schedule formulated as
\begin{equation}\label{eq:omega_k}
\omega_k = \frac{1}{10000^{2k/d_{\mathrm{model}}}}.
\end{equation}

The column encoding vector components $[\bm{E}_{\mathrm{col}}(j)]_{2k}$ and $[\bm{E}_{\mathrm{col}}(j)]_{2k+1}$ are derived analogously for the vertical index $j$ utilizing the identical frequency schedule $\omega_k$, ensuring that the structural grid dimensions map uniquely without losing spatial topology info.

To capture the high-dimensional non-linear channel correlations across the visible ports, the augmented latent matrix $\bm{X}_{\mathrm{p}}'$ is routed through a series of specialized encoder blocks. In order to enhance feature extraction and boost resilience against deep fading noise, each constituent block is designed as an interleaved micro-topology combining multi-head self-attention ($\mathrm{MSA}$) modules, sparse Mixture-of-Experts ($\mathrm{MoE}$) layers, and standard multi-layer perceptron feed-forward networks ($\mathrm{FFN}$). Layer normalization ($\mathrm{LN}$) and residual connections encompass each operation to ensure numerical stability. Within a given block, the input features are first contextualized by the primary attention pass according to
\begin{equation}\label{eq:msa_first}
\bm{X}_{\mathrm{p}}'' = \mathrm{LN}\left(\bm{X}_{\mathrm{p}}' + \mathrm{MSA}\left(\bm{X}_{\mathrm{p}}'\right)\right).
\end{equation}

Subsequently, the intermediate features $\bm{X}_{\mathrm{p}}''$ are injected into the $\mathrm{MoE}$ layer. This architecture comprises $E$ parallel expert networks $\{\mathcal{E}_1, \dots, \mathcal{E}_E\}$ and a trainable gating network $\mathcal{G}$, where the total number of experts is configured as $E = 4$. Each individual expert $\mathcal{E}_j$ executes a non-linear coordinate mapping parameterized as
\begin{equation}\label{eq:expert_def}
\mathcal{E}_j\left(\bm{X}_{\mathrm{p}}''\right) = \sigma\left(\bm{X}_{\mathrm{p}}'' \bm{W}_{\mathrm{e}1}^{(j)}\right)\bm{W}_{\mathrm{e}2}^{(j)},
\end{equation}
where $\sigma(\cdot)$ denotes the Gaussian Error Linear Unit ($\mathrm{GELU}$) activation function, while $\bm{W}_{\mathrm{e}1}^{(j)} \in \mathbb{R}^{d_{\mathrm{model}} \times d_{\mathrm{h}}}$ and $\bm{W}_{\mathrm{e}2}^{(j)} \in \mathbb{R}^{d_{\mathrm{h}} \times d_{\mathrm{model}}}$ represent the trainable expert weight layers with a hidden dimension expansion profile of $d_{\mathrm{h}} = 4d_{\mathrm{model}} = 768$. Concurrently, the gating network evaluates the routing affinity vector $\bm{g} \in \mathbb{R}^{N_{\mathrm{o}} \times E}$ using a linear projection layer combined with a softmax activation operator as
\begin{equation}\label{eq:gating_prob}
\bm{g} = \mathrm{softmax}\left(\bm{X}_{\mathrm{p}}'' \bm{W}_{\mathrm{g}} + \bm{1}\bm{b}_{\mathrm{g}}^{\mathrm{T}}\right).
\end{equation}
where $\bm{W}_{\mathrm{g}} \in \mathbb{R}^{d_{\mathrm{model}} \times E}$ is the learnable gating weight matrix, $\bm{b}_{\mathrm{g}} \in \mathbb{R}^{E \times 1}$ is the gating bias vector, and $\bm{1} \in \mathbb{R}^{N_{\mathrm{o}} \times 1}$ denotes a vector of all ones. To maintain a lightweight computational footprint, a top-$K$ sparse activation mechanism filters the gating scores such that only $K = 2$ experts are activated per port token:
\begin{equation}\label{eq:topk_selection}
\{s_k\}_{k=1}^K = \mathrm{TopK}(\bm{g}, K).
\end{equation}

The definitive output of the $\mathrm{MoE}$ layer aggregates the weighted expert outputs combined with a dropout regularization operator of probability $p = 0.1$, which is given by
\begin{equation}\label{eq:moe_aggregation}
\bm{X}_{\mathrm{p}}''' = \mathrm{Dropout}\left(\sum_{k=1}^K s_k \cdot \mathcal{E}_{l_k}\left(\bm{X}_{\mathrm{p}}''\right)\right),
\end{equation}
where $l_k$ represents the exact index of the $k$-th selected expert network.

Following the expert aggregation layer, the tokens are further propagated through a secondary $\mathrm{MSA}$ pass and a standard multi-layer perceptron block to finalize the intra-block representation stream:
\begin{align}
\bm{X}_{\mathrm{p}}'''' &= \mathrm{LN}\left(\bm{X}_{\mathrm{p}}''' + \mathrm{MSA}\left(\bm{X}_{\mathrm{p}}'''\right)\right), \label{eq:msa_second} \\
\bm{X}_{\mathrm{p}}''''' &= \mathrm{LN}\left(\bm{X}_{\mathrm{p}}'''' + \mathrm{FFN}\left(\bm{X}_{\mathrm{p}}''''\right)\right). \label{eq:ffn_final}
\end{align}
This comprehensive sequence of block operations is repeated over a depth of $N_{\mathrm{blocks}}/2$ times, yielding the definitive encoder latent feature matrix $\bm{X}_{\mathrm{enc}} \in \mathbb{R}^{N_{\mathrm{o}} \times d_{\mathrm{model}}}$. The output representation matrix $\bm{X}_{\mathrm{enc}}$ effectively compresses the non-linear spatial dependencies of the fluid antenna channel while maintaining strict physical spatial order, acting as the primary anchor sequence before entering the lightweight decoder framework.

\begin{figure}[!t]
\centering\includegraphics[width=0.5\textwidth]{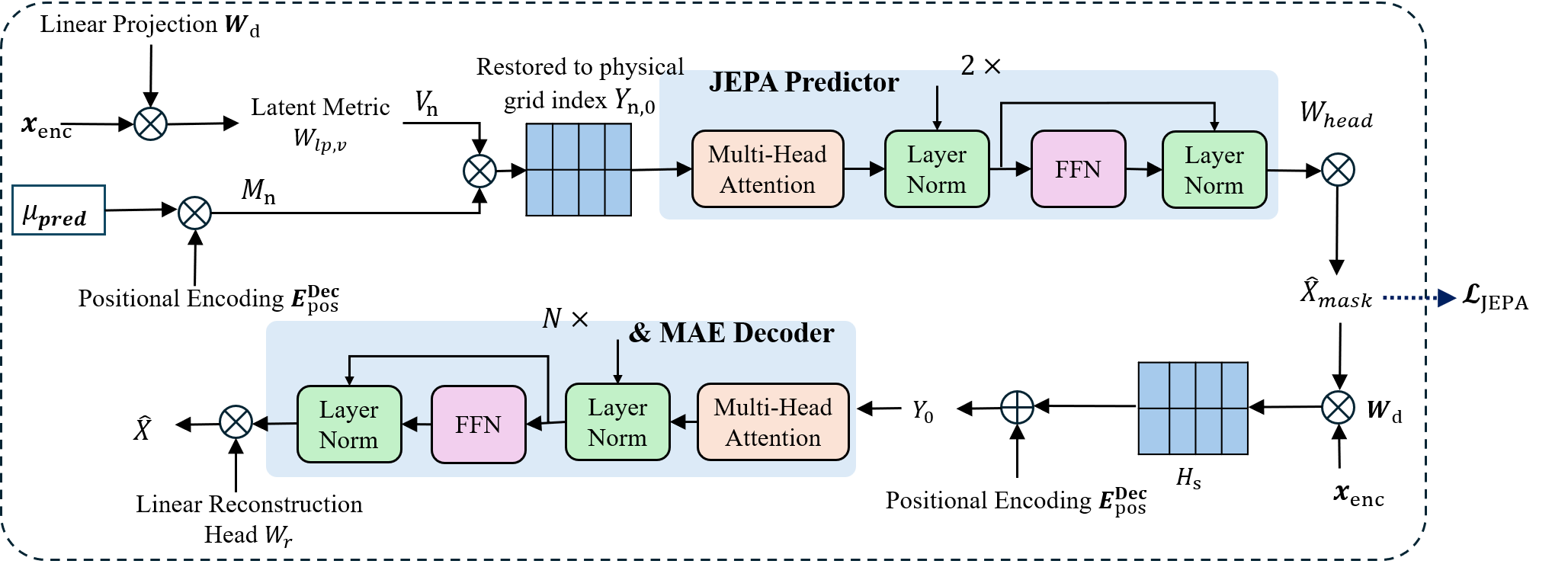}
\caption{The architecture of the JEPA latent predictor and decoder.}
\label{fig:JEPA_decoder}
\end{figure}

\subsubsection{JEPA Latent Predictor}
To optimize the channel foundation model without relying exclusively on high-dimensional raw signal reconstruction, the pre-training framework incorporates a JEPA branch, which is illustrated in Fig. \ref{fig:JEPA_decoder}. This predictive mechanism directs the encoder to learn structurally robust channel representations by forecasting the latent embeddings of the unobserved segments within a highly compressed feature space.

The target latent representations are generated via a parallel full-view pass through the identical encoder backbone formulated in Section B. Let $\bm{X}_n \in \mathbb{R}^{2M \times N_{\mathrm{X}} \times N_{\mathrm{Y}}}$ denote the complete, unmasked standardized channel state information tensor of the $n$-th sample. During this target generation phase, the spatial masking operator is deactivated by enforcing a mask ratio of $0.0$, thereby ensuring that all $K_{\mathrm{p}} = 1024$ patch segments remain visible. By propagating this uncorrupted full-view sequence through the linear projection layer and the cascaded Transformer blocks, a full-view latent feature matrix $\bm{X}_{\mathrm{full}} \in \mathbb{R}^{K_{\mathrm{p}} \times d_{\mathrm{model}}}$ is achieved. To isolate the target characteristics corresponding to the unobserved wireless ports, the true masked latent matrix $\bm{X}_{\mathrm{mask},\mathrm{true}} \in \mathbb{R}^{K_{\mathrm{mask}} \times d_{\mathrm{model}}}$ is constructed by extracting the row configurations indexed by the complement masked subset $\mathcal{I}_{\mathrm{mask},n}$, where $K_{\mathrm{mask}} = K_{\mathrm{p}} - N_{\mathrm{o}} = 768$. The extraction is mathematically expressed as
\begin{equation}\label{eq:true_latent_extract}
[\bm{X}_{\mathrm{mask},\mathrm{true}}]_{k,:} = \left[ \bm{X}_{\mathrm{full}} \right]_{\mathcal{I}_{\mathrm{mask},n}(k),:}.
\end{equation}
where $1 \le k \le K_{\mathrm{mask}}$. To minimize memory footprints and satisfy the architectural conditions of the JEPA paradigm, the gradient propagation stream through this full-view target pass is deactivated, yielding a detached target representation matrix.
Concurrently, the latent predictor network synthesizes these unobserved patch features within the embedding domain, relying solely on the visible latent feature matrix $\bm{X}_{\mathrm{enc}} \in \mathbb{R}^{N_{\mathrm{o}} \times d_{\mathrm{model}}}$ obtained in \eqref{eq:ffn_final}. The visible token sequence is first projected into a localized predictive space via a trainable linear layer parameterized by the weight matrix $\bm{W}_{\mathrm{lp},\mathrm{v}} \in \mathbb{R}^{d_{\mathrm{model}} \times d_{\mathrm{model}}}$, producing the projected visible matrix $\bm{V}_n \in \mathbb{R}^{N_{\mathrm{o}} \times d_{\mathrm{model}}}$ as
\begin{equation}\label{eq:visible_pred_proj}
\bm{V}_n = \bm{X}_{\mathrm{enc}} \bm{W}_{\mathrm{lp},\mathrm{v}}.
\end{equation}

For the unobserved segments, a learnable mask token vector $\bm{\mu}_{\mathrm{pred}} \in \mathbb{R}^{1 \times d_{\mathrm{model}}}$ acts as a uniform spatial placeholder for each missing patch. To maintain physical coordinate awareness within the predictor layer, a trainable predictor positional embedding matrix $\bm{E}_{\mathrm{pred},\mathrm{pos}} \in \mathbb{R}^{K_{\mathrm{mask}} \times d_{\mathrm{model}}}$ is superposed element-wise onto these placeholders, defining the masked token matrix $\bm{M}_n \in \mathbb{R}^{K_{\mathrm{mask}} \times d_{\mathrm{model}}}$ via its components:
\begin{equation}\label{eq:masked_placeholder}
[\bm{M}_n]_{k,:} = \bm{\mu}_{\mathrm{pred}} + [\bm{E}_{\mathrm{pred},\mathrm{pos}}]_{k,:}.
\end{equation}

The rows of the projected visible feature matrix $\bm{V}_n$ and the spatially augmented placeholder matrix $\bm{M}_n$ are subsequently recombined and restored to their original physical grid index configurations. This sorting process yields the unified initial predictor input matrix $\bm{Y}_{n,0} \in \mathbb{R}^{K_{\mathrm{p}} \times d_{\mathrm{model}}}$, which strictly respects the spatial grid topology of the 2D FAS layout.

The recombined matrix $\bm{Y}_{n,0}$ is propagated through $N_{\mathrm{pred\_blocks}} = 2$ lightweight Transformer blocks to model global contextual dependencies across the complete grid aperture. Each block contains an $\mathrm{MSA}$ pass and an $\mathrm{FFN}$ layer configured with a compact hidden expansion profile. Let $\bm{Y}_{n,N_{\mathrm{pred\_blocks}}}$ designate the definitive feature matrix generated by the terminal predictor block. Following a layer normalization operation, the contextualized features at the masked index locations $\mathcal{I}_{\mathrm{mask},n}$ are isolated and mapped through a linear prediction head with weight matrix $\bm{W}_{\mathrm{head}} \in \mathbb{R}^{d_{\mathrm{model}} \times d_{\mathrm{model}}}$. The final predicted masked latent matrix $\bm{\hat{X}}_{\mathrm{mask}} \in \mathbb{R}^{K_{\mathrm{mask}} \times d_{\mathrm{model}}}$ is formally modeled as
\begin{equation}\label{eq:predicted_masked_latent}
\begin{split}
[\bm{\hat{X}}_{\mathrm{mask}}]_{k,:} = \mathrm{LN}\bigg( \big[ \bm{Y}_{n,N_{\mathrm{pred\_blocks}}} \big]_{\mathcal{I}_{\mathrm{mask},n}(k),:} \bigg) \\
\times \bm{W}_{\mathrm{head}}.
\end{split}
\end{equation}

The optimization objective of the JEPA branch minimizes the mean squared error between the forecasted embeddings and the true target representations within the latent domain, penalizing structural discrepancies without incurring pixel-level high-dimensional decoding overhead. The self-supervised JEPA loss function $\mathcal{L}_{\mathrm{JEPA}}$ is defined as
\begin{equation}\label{eq:jepa_loss}
\mathcal{L}_{\mathrm{JEPA}} = \frac{1}{K_{\mathrm{mask}}} \left\| \bm{\hat{X}}_{\mathrm{mask}} - \bm{X}_{\mathrm{mask},\mathrm{true}} \right\|_{\mathrm{F}}^2.
\end{equation}

By computing the loss strictly within the compact embedding space, the model bypasses the computational burdens of raw coefficient generation, forcing the encoder to learn stable, high-level geometric patterns of the multi-path propagation medium.

\subsubsection{Decoder and Channel Extrapolation}
To project the latent representation stream back into the physical domains and achieve full-grid channel state information reconstruction, a lightweight masked autoencoder (MAE) decoder architecture is deployed. As illustrated in Fig. \ref{fig:JEPA_decoder}, the decoder serves as the optimization head for Task A, synthesizing the visible features from the encoder and the forecasted features from the JEPA predictor to counteract localized spatial field-of-view blockages.

Let $\bm{Z}_{\mathrm{enc}} \in \mathbb{R}^{(1 + N_{\mathrm{o}}) \times d_{\mathrm{m}}}$ denote the definitive output sequence generated by the masked-view encoder pass, where $N_{\mathrm{o}}$ represents the number of observable ports and $d_{\mathrm{m}}$ is the encoder hidden dimension. We isolate the visible port token matrix $\bm{X}_{\mathrm{enc}} \in \mathbb{R}^{N_{\mathrm{o}} \times d_{\mathrm{m}}}$ from the row space of $\bm{Z}_{\mathrm{enc}}$ spanning indices $2$ to $1+N_{\mathrm{o}}$, and let $\bm{z}_{\mathrm{cls}} \in \mathbb{R}^{1 \times d_{\mathrm{m}}}$ be the companion encoder classification token. Concurrently, let $\bm{\hat{X}}_{\mathrm{mask}} \in \mathbb{R}^{N_{\mathrm{m}} \times d_{\mathrm{m}}}$ represent the predicted latent matrix for the $N_{\mathrm{m}} = N_{\mathrm{S}} - N_{\mathrm{o}}$ masked ports. The structural attributes are first projected into a lower-dimensional space via a trainable linear layer characterized by the weight matrix $\bm{W}_{\mathrm{d}} \in \mathbb{R}^{d_{\mathrm{m}} \times d_{\mathrm{d}}}$, where $d_{\mathrm{d}} = 64$ defines the compact decoder hidden profile. The dimension-reduced token blocks are mathematically formulated as
\begin{align}
\bm{H}_{\mathrm{o}} &= \bm{X}_{\mathrm{enc}} \bm{W}_{\mathrm{d}}, \label{eq:dec_proj_vis} \\
\bm{H}_{\mathrm{m}} &= \bm{\hat{X}}_{\mathrm{mask}} \bm{W}_{\mathrm{d}},\label{eq:dec_proj_mask}
\end{align}
where $\bm{H}_{\mathrm{o}} \in \mathbb{R}^{N_{\mathrm{o}} \times d_{\mathrm{d}}}$ and $\bm{H}_{\mathrm{m}} \in \mathbb{R}^{N_{\mathrm{m}} \times d_{\mathrm{d}}}$.

To restore the strict spatial coordinate relationships of the 2D FAS grid apertures, the projected visible and predicted masked sequences are integrated. The matrices are first concatenated along the spatial axis to construct an intermediate grid set $\bm{H}_{\mathrm{c}} = [ \bm{H}_{\mathrm{o}}^{\mathrm{T}}, \bm{H}_{\mathrm{m}}^{\mathrm{T}} ]^{\mathrm{T}} \in \mathbb{R}^{N_{\mathrm{S}} \times d_{\mathrm{d}}}$. A spatial permutation operator parameterized by the deterministic binary matching matrix $\bm{\Pi} \in \{0, 1\}^{N_{\mathrm{S}} \times N_{\mathrm{S}}}$ is applied sample-wise to map the sequence back into its planar topology layout, yielding the ordered matrix $\bm{H}_{\mathrm{s}} = \bm{\Pi} \bm{H}_{\mathrm{c}}$. Prepending the decoder classification token $\bm{h}_{\mathrm{cls}} = \bm{z}_{\mathrm{cls}} \bm{W}_{\mathrm{d}}$ at index $0$ and superposing a fixed two-dimensional decoder positional embedding matrix $\bm{E}_{\mathrm{d}} \in \mathbb{R}^{(1 + N_{\mathrm{S}}) \times d_{\mathrm{d}}}$, the collective initial feature sequence $\bm{Y}_{0} \in \mathbb{R}^{(1 + N_{\mathrm{S}}) \times d_{\mathrm{d}}}$ is achieved as
\begin{equation}\label{eq:decoder_initial_concat}
\bm{Y}_{0} = \begin{bmatrix} \bm{h}_{\mathrm{cls}} & \bm{H}_{\mathrm{s}} \end{bmatrix} + \bm{E}_{\mathrm{d}}.
\end{equation}

The spatially aligned representation sequence $\bm{Y}_{0}$ is propagated through a deep vision backbone composed of $N_{\mathrm{d}} = 4$ identical cascaded Transformer decoding blocks. Let $\bm{Y}_{f}$ designate the output matrix of the $f$-th block layer, where $1 \le f \le N_{\mathrm{d}}$. Utilizing layer normalization and residual connections, the intra-block representation updates are governed by the following expressions:
\begin{align}
\bm{Y}_{f}' &= \bm{Y}_{f-1} + \mathrm{MSA}\left( \mathrm{LN}\left( \bm{Y}_{f-1} \right) \right), \label{eq:msa_dec_block} \\
\bm{Y}_{f} &= \bm{Y}_{f}' + \mathrm{FFN}\left( \mathrm{LN}\left( \bm{Y}_{f}' \right) \right). \label{eq:ffn_dec_block}
\end{align}

Upon execution of the final block layer, the sequence is normalized, and the classification token at index $0$ is stripped to isolate the decoded spatial matrix $\bm{H}_{\mathrm{out}} \in \mathbb{R}^{N_{\mathrm{S}} \times d_{\mathrm{d}}}$. A linear reconstruction head characterized by the layer weight matrix $\bm{W}_{\mathrm{r}} \in \mathbb{R}^{d_{\mathrm{d}} \times 2M}$ maps the latent representation directly into the real coefficient space, yielding the reconstructed full-grid standardized channel tensor $\bm{\hat{X}} = \bm{H}_{\mathrm{out}} \bm{W}_{\mathrm{r}} \in \mathbb{R}^{N_{\mathrm{S}} \times 2M}$.

The fundamental optimization objective of Task A is to enforce pixel-level reconstruction fidelity specifically over the unobserved channel regions. Let $\bm{X}_n \in \mathbb{R}^{N_{\mathrm{S}} \times 2M}$ and $\bm{\hat{X}}_n \in \mathbb{R}^{N_{\mathrm{S}} \times 2M}$ denote the ground-truth clean channel tensor and the reconstructed tensor for the $n$-th sample, respectively. Let $\bm{x}_{n,k}^{\mathrm{T}}$ and $\bm{\hat{x}}_{n,k}^{\mathrm{T}}$ represent the $k$-th rows of these matrices, corresponding to the $2M$-dimensional real-valued CSI vector at port index $k$. Let $m_{n,k} \in \{0, 1\}$ define the mask indicator variable, where $m_{n,k} = 1$ if port $k$ is masked during training and $m_{n,k} = 0$ if it is observable. Accumulating the errors over a training batch size of $B$, the self-supervised MAE reconstruction loss function $\mathcal{L}_{\mathrm{MAE}}$ is formulated as
\begin{equation}\label{eq:mae_loss_final}
\mathcal{L}_{\mathrm{MAE}} = \frac{\sum_{n=1}^{B} \sum_{k=1}^{N_{\mathrm{S}}} m_{n,k} \left\| \bm{\hat{x}}_{n,k} - \bm{x}_{n,k} \right\|_2^2}{\sum_{n=1}^{B} \sum_{k=1}^{N_{\mathrm{S}}} m_{n,k} + \epsilon},
\end{equation}
where $\epsilon = 10^{-8}$ represents the numerical protection regularizer. 

\subsubsection{Sliced Isotropic Gaussian Regularization}
To prevent representation collapse within the latent embedding space and safeguard the foundation network against mapping diverse channel state signatures into a trivial invariant manifold, a Sliced Isotropic Gaussian Regularization (SIGReg) mechanism is integrated. This regularizer guides the distribution of the generated latent embeddings toward an isotropic standard normal distribution by minimizing the sliced characteristic function distance.

Let $\bm{V} \in \mathbb{R}^{N_{\mathrm{e}} \times d_{\mathrm{m}}}$ denote a sub-sampled matrix of latent feature vectors extracted from either the full-view or visible token stream during pre-training, where $d_{\mathrm{m}}$ represents the embedding dimension and $N_{\mathrm{e}}$ is the regularizer evaluation batch size bounded by a maximum threshold of $2048$. To evaluate the multidimensional distribution properties efficiently, the high-dimensional features are projected onto a group of uniform random directions. Let $\bm{A} \in \mathbb{R}^{N_{\mathrm{s}} \times d_{\mathrm{m}}}$ represent the slice direction matrix containing $N_{\mathrm{s}} = 128$ distinct projection vectors. Each individual row vector $\bm{a}_s^{\mathrm{T}}$ of the slice matrix $\bm{A}$ is normalized to unit length according to
\begin{equation}\label{eq:slice_normalization}
\bm{a}_s = \frac{\bm{\tilde{a}}_s}{\left\| \bm{\tilde{a}}_s \right\|_2 + \zeta},
\end{equation}where $\bm{\tilde{a}}_s \in \mathbb{R}^{d_{\mathrm{m}} \times 1}$ is a random vector drawn from an independent and identically distributed standard normal distribution $\mathcal{N}(\bm{0}, \bm{I})$, $1 \le s \le N_{\mathrm{s}}$, and $\zeta = 10^{-6}$ represents a minor numerical stabilization parameter. The projected representations are captured via the matrix multiplication $\bm{P} = \bm{V}\bm{A}^{\mathrm{T}} \in \mathbb{R}^{N_{\mathrm{e}} \times N_{\mathrm{s}}}$, where the scalar element $p_{n,s} = [\bm{P}]_{n,s}$ represents the one-dimensional coordinate projection of the $n$-th token embedding along the $s$-th random slice direction.

The SIGReg framework enforces a statistical penalty based on the Cramer-von Mises distance between the empirical characteristic function (ECF) of these projected coordinates and the theoretical characteristic function of a standard Gaussian distribution. For a given slice direction $s$ and a real-valued evaluation frequency parameter $t \in \mathbb{R}$, the theoretical Gaussian characteristic function is designated as $\phi(t) = \exp(-0.5 t^2)$, while the corresponding ECF computed over the sampled token distribution is defined as
\begin{equation}\label{eq:empirical_cf}
\hat{\phi}_s(t) = \frac{1}{N_{\mathrm{e}}} \sum_{n=1}^{N_{\mathrm{e}}} \exp(\mathrm{j} t p_{n,s}),
\end{equation}where $\mathrm{j} = \sqrt{-1}$ represents the imaginary unit. To approximate the integral distance over a continuous spectral span, a uniform grid consisting of $N_{\mathrm{t}} = 17$ points is defined across the domain $[\tau_{\mathrm{min}}, \tau_{\mathrm{max}}] = [-5.0, 5.0]$. Let $\mathcal{T} = \{t_1, t_2, \dots, t_{N_{\mathrm{t}}}\}$ denote this uniform evaluation array, where the consistent grid interval is expressed as $\Delta t = ( \tau_{\mathrm{max}} - \tau_{\mathrm{min}} ) / (N_{\mathrm{t}} - 1)$.

By evaluating the discrepancy at each discrete grid frequency via a trapezoidal integration rule, the localized distribution misalignment for the $s$-th slice direction is quantified by the scalar integral value $I_s$, which is mathematically formulated as
\begin{equation}\label{eq:trapezoid_integral}
\begin{split}
I_s = \Delta t \Bigg[ \frac{1}{2} \left| \hat{\phi}_s(t_1) - \phi(t_1) \right|^2 + \frac{1}{2} \left| \hat{\phi}_s(t_{N_{\mathrm{t}}}) - \phi(t_{N_{\mathrm{t}}}) \right|^2 \\
+ \sum_{k=2}^{N_{\mathrm{t}}-1} \left| \hat{\phi}_s(t_k) - \phi(t_k) \right|^2 \Bigg].
\end{split}
\end{equation}

The definitive self-supervised statistical regularization loss function $\mathcal{L}_{\mathrm{SIGReg}}$ is obtained by averaging the computed trapezoidal grid integrals across all random slice directions, yielding
\begin{equation}\label{eq:sigreg_loss_final}
\mathcal{L}_{\mathrm{SIGReg}} = \frac{1}{N_{\mathrm{s}}} \sum_{s=1}^{N_{\mathrm{s}}} I_s.
\end{equation}
By computing this objective within a non-parametric characteristic domain, the SIGReg loss isolates and smooths out distribution anomalies across the latent spaces without requiring explicit density reconstruction, thereby stabilizing the underlying backbone against representation collapse.

\subsection{Downstream wireless positioning}

To evaluate the capability of the learned channel foundation representations in capturing fine-grained spatial propagation geometries, the pre-trained encoder backbone is adapted to the downstream task of high-precision wireless localization. This task seeks to map the complex multi-path spatial fingerprints recorded across the full aperture of the two-dimensional FAS directly onto the three-dimensional physical coordinates of the UE.

During the downstream fine-tuning phase, the pre-trainind encoder and JEPA encoder are utilized. The general features $\bm{\hat{X}}_{\mathrm{mask}}$ generated through the masked-view encoder pass is passed to a global average pooling (GAP) operation is performed across the entire physical patch token sequence to aggregate the multi-path geometric components. The pooled spatial representation vector $\bm{x}_{\mathrm{gap}} \in \mathbb{R}^{1 \times d_{\mathrm{m}}}$ is formally defined as
\begin{equation}\label{eq:gap_operation}
\bm{x}_{\mathrm{gap}} = \frac{1}{K_{\mathrm{p}}} \sum_{k=2}^{1+K_{\mathrm{p}}} [\bm{\hat{X}}_{\mathrm{mask}}]_{k,:}.
\end{equation}

The pooled fingerprint vector $\bm{x}_{\mathrm{gap}}$ is then propagated through a dedicated position regression head consisting of an expanded and deep multi-layer perceptron (MLP) structure. The regression framework sequentially maps the hidden features through hidden dimensions of $512$ and $256$ before projecting onto the three-dimensional coordinate space. Utilizing the Gaussian error linear unit ($\mathrm{GELU}$) non-linear operator $\sigma(\cdot)$, the intermediate coordinate mapping layers are formulated as
\begin{align}
\bm{h}_1 &= \sigma\left( \bm{x}_{\mathrm{gap}} \bm{W}_{\mathrm{l1}} + \bm{b}_{\mathrm{l1}}^{\mathrm{T}} \right), \label{eq:mlp_layer1} \\
\bm{h}_2 &= \sigma\left( \bm{h}_1 \bm{W}_{\mathrm{l2}} + \bm{b}_{\mathrm{l2}}^{\mathrm{T}} \right). \label{eq:mlp_layer2}
\end{align}
where $\bm{W}_{\mathrm{l1}} \in \mathbb{R}^{d_{\mathrm{m}} \times 512}$ and $\bm{W}_{\mathrm{l2}} \in \mathbb{R}^{512 \times 256}$ denote the trainable layer weight matrices, while $\bm{b}_{\mathrm{l1}} \in \mathbb{R}^{512 \times 1}$ and $\bm{b}_{\mathrm{l2}} \in \mathbb{R}^{256 \times 1}$ represent the corresponding bias vectors. The final estimated relative coordinate vector $\bm{\hat{y}}_n \in \mathbb{R}^{1 \times 3}$ is generated via a terminal linear projection layer as
\begin{equation}\label{eq:mlp_output}
\bm{\hat{y}}_n = \bm{h}_2 \bm{W}_{\mathrm{l3}} + \bm{b}_{\mathrm{l3}}^{\mathrm{T}}.
\end{equation}
where $\bm{W}_{\mathrm{l3}} \in \mathbb{R}^{256 \times 3}$ and $\bm{b}_{\mathrm{l3}} \in \mathbb{R}^{3 \times 1}$ define the terminal layer weight and bias parameters, respectively.

Let $\bm{y}_{n,\mathrm{pos}} = [x_n, y_n, z_n] \in \mathbb{R}^{1 \times 3}$ denote the absolute ground-truth physical coordinate vector of the UE in the physical simulation area. To safeguard numerical gradient stability during the backpropagation sequence, the absolute positions are scaled down by a constant spatial normalization factor of $350.0\,\mathrm{m}$, yielding the standardized target relative coordinate vector $\bm{y}_n = \bm{y}_{n,\mathrm{pos}} / 350.0$. The downstream regression head and the underlying encoder backbone are optimized jointly over a training batch size of $B$ by minimizing the mean absolute error ($\mathcal{L}_1$ loss) function, which is expressed as
\begin{equation}\label{eq:l1_localization_loss}
\mathcal{L}_{\mathrm{loc}} = \frac{1}{B} \sum_{n=1}^{B} \left\| \bm{\hat{y}}_n - \bm{y}_n \right\|_1.
\end{equation}
\section{Simulation results and analysis}
\label{sec:simulation_results}
\subsection{Simulation Settings}
\begin{figure}[!t]
\centering\includegraphics[width=0.5\textwidth]{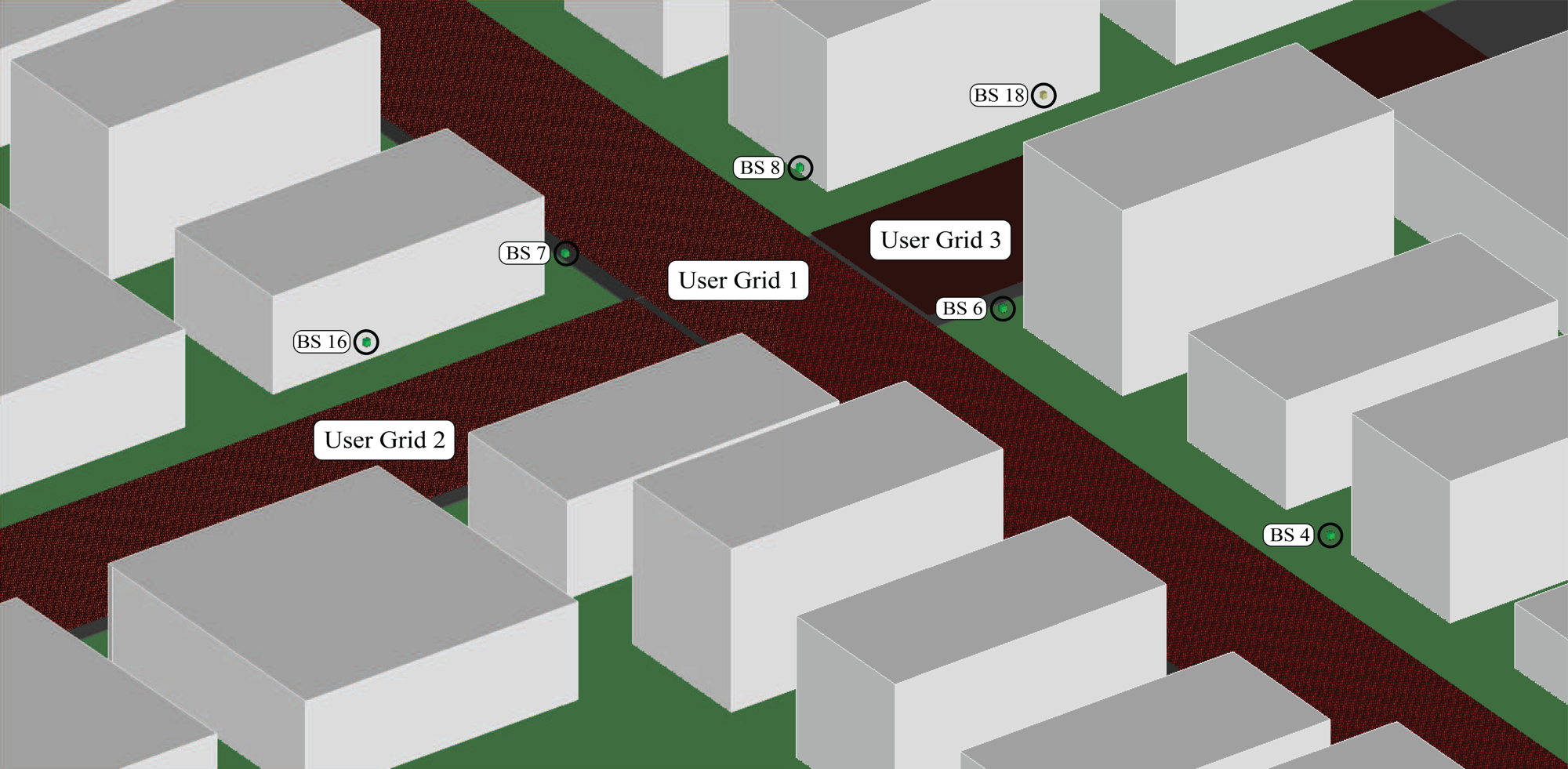}
\caption{The layout of the simulation scenario.}
\label{fig:scenarios}
\end{figure}
To thoroughly evaluate the performance of the proposed joint embedding predictive architecture-based channel foundation model (JEPA-CFM), we conducted extensive numerical simulations. The propagation environment is modeled using the site-specific ray-tracing DeepMIMO O1\_28 scenario, which represents a typical urban outdoor deployment (Fig.~\ref{fig:scenarios}). The base station is equipped with a uniform linear array comprising $  M = 8  $ antennas spaced at $  d_{\mathrm{BS}} = 0.5\lambda  $. At the user equipment, a two-dimensional fluid antenna system spans a continuous rectangular aperture that is discretized into an $  N_{\mathrm{X}} \times N_{\mathrm{Y}} = 64 \times 64  $ grid, yielding a total of $  N_{\mathrm{S}} = 4096  $ switchable ports with uniform spacing $  d_{\mathrm{X}} = d_{\mathrm{Y}} = 0.1\lambda  $. The channel responses are synthesized from $  L = 5  $ dominant propagation paths.

A dataset of 30,000 independent channel realizations was generated via ray-tracing and divided into blocks of 5,000 samples to facilitate efficient storage and processing. The full dataset was partitioned into training and testing sets using a 9:1 ratio, resulting in 27,000 training samples and 3,000 independent test samples. To assess the model across the two primary tasks, we established the following benchmarks:
\begin{enumerate}
\item For the spatial channel extrapolation task (Task A), the proposed JEPA-CFM framework is compared against a conventional pure masked autoencoder (Pure MAE) baseline to demonstrate the benefits of predictive latent-space representation learning under severe aperture blockages.
\item For the downstream wireless localization task (Task B), the coordinate regression head attached to the pre-trained encoder is evaluated against a traditional mathematical fingerprint-matching algorithm to demonstrate the high-resolution geometric awareness captured by the learned embeddings under complex multipath conditions.
\end{enumerate}
\subsection{Model Training and Loss Formulations}
\begin{figure}[!t]
\centering\includegraphics[width=0.5\textwidth]{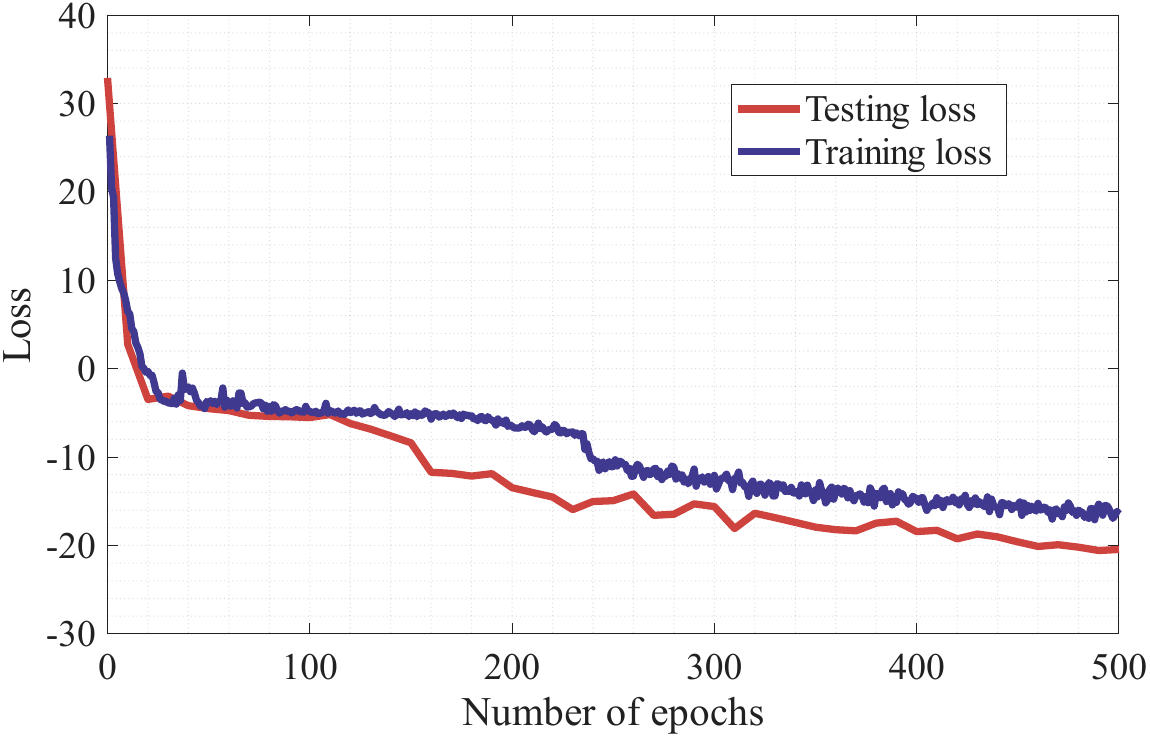}
\caption{The training and testing loss of the proposed JEPA-CFM.}
\label{fig:loss}
\end{figure}
The JEPA-CFM framework is trained in two distinct phases: self-supervised multi-objective pre-training followed by task-specific fine-tuning. Pre-training was performed on a distributed multi-GPU system using two nodes. The process ran for 700 epochs with the AdamW optimizer and a mini-batch size of $  B_{\mathrm{pre}} = 128  $. A random spatial mask with ratio $  M_{\mathrm{r}} = 0.75  $ was applied throughout, limiting the encoder to only 25\% of the total fluid antenna ports. The overall pre-training objective minimizes the combined loss 
\begin{equation}
\mathcal{L}_{\mathrm{pre}} = \gamma \mathcal{L}_{\mathrm{MAE}} + \lambda_{\mathrm{JEPA}} \mathcal{L}_{\mathrm{JEPA}} + \lambda_{\mathrm{SIGReg}} \mathcal{L}_{\mathrm{SIGReg}},    
\end{equation}where the scaling factor $  \gamma = 10000.0  $ balances the pixel-level reconstruction term against the latent constraints, while $  \lambda_{\mathrm{JEPA}} = 0.2  $ and $  \lambda_{\mathrm{SIGReg}} = 0.02  $ weight the JEPA prediction and regularization terms, respectively. The evolution of training and testing loss is shown in Fig.~\ref{fig:loss}.
To quantify extrapolation accuracy on the test set, we employ the normalized mean squared error (NMSE) computed exclusively over the unobserved (masked) regions:
\begin{equation}
    \mathrm{NMSE} = \frac{\sum_{n=1}^{B} \sum_{k=1}^{N_{\mathrm{S}}} m_{n,k} \left\| \bm{\hat{x}}_{n,k} - \bm{x}_{n,k} \right\|_2^2}{\sum_{n=1}^{B} \sum_{k=1}^{N_{\mathrm{S}}} m_{n,k} \left\| \bm{x}_{n,k} \right\|_2^2 + \epsilon}.
\end{equation}

This metric ensures that evaluation focuses solely on the model’s ability to infer hidden ports, thereby validating its universal extrapolation capability. After pre-training, the base encoder parameters are frozen for the downstream positioning task. Fine-tuning is performed on four computing nodes with a mini-batch size of $  B_{\mathrm{ft}} = 36  $, a learning rate of $2.0 \times 10^{-4}$, and no weight decay (the mask ratio remains $  M_{\mathrm{r}} = 0.75  $). The position regression head is optimized by minimizing the mean absolute error $  \mathcal{L}_{\mathrm{loc}}  $ defined in \eqref{eq:l1_localization_loss}. All simulation parameters and training hyper-parameters are summarized in Table~\ref{tab:simulation_settings}.
\begin{table}[!t]
\renewcommand{\arraystretch}{1.2}
\caption{Simulation settings and training hyper-parameters}
\label{tab:simulation_settings}
\centering
\begin{tabular}{p{5.5cm}c}
\hline
\textbf{Simulation Parameters} & \textbf{Values} \\
\hline
Simulation scenario & DeepMIMO O1\_28 \\
BS antenna array & 8-element ULA \\
BS antenna spacing $d_{\mathrm{BS}}$ & $0.5\lambda$ \\
FAS port $N_{\mathrm{X}} \times N_{\mathrm{Y}}$ & $64 \times 64$ \\
Total selectable physical ports $N_{\mathrm{S}}$ & 4096 \\
FAS port spacing $d_{\mathrm{X}}, d_{\mathrm{Y}}$ & $0.1\lambda$ \\
Number of dominant propagation paths $L$ & 5 \\
Total simulated CSI samples & 30000 \\
Data block chunk processing size & 5000 \\
Dataset partition training/testing split ratio & 9:1 \\
\hline
\textbf{Training Hyper-parameters} & \textbf{Values} \\
\hline
Pre-training Mini-batch Size $B_{\mathrm{pre}}$ & 128 \\
Total number of pre-training epochs & 500 \\
Base learning rate parameter $\eta_{\mathrm{pre}}$ & $5.0 \times 10^{-4}$ \\
Weight decay coefficient & 0.05 \\
Random spatial mask ratio variable $M_{\mathrm{r}}$ & 0.75 \\
Weight of JEPA loss $\lambda_{\mathrm{JEPA}}$ & 0.2 \\
Weight of SIGReg loss $\lambda_{\mathrm{SIGReg}}$ & 0.02 \\
Weight of MAE loss $\gamma$ & 10000.0 \\
Deep-fading regularizer constant $\epsilon$ & $10^{-8}$ \\
\hline
\textbf{Downstream positioning training} & \textbf{Values} \\
\hline
Fine-tuning mini-batch Size $B_{\mathrm{ft}}$ & 36 \\
Fine-tuning initial learning rate $\eta_{\mathrm{ft}}$ & $2.0 \times 10^{-4}$ \\
Fine-tuning weight decay parameter & 0.0 \\
Fine-tuning operational mask ratio $M_{\mathrm{r}}$ & 0.75 \\
Physical coordinate normalization factor & $350.0\,\mathrm{m}$ \\
\hline
\end{tabular}
\end{table}
\subsubsection{Channel extrapolation}
\begin{figure}[!t]
\centering\includegraphics[width=0.5\textwidth]{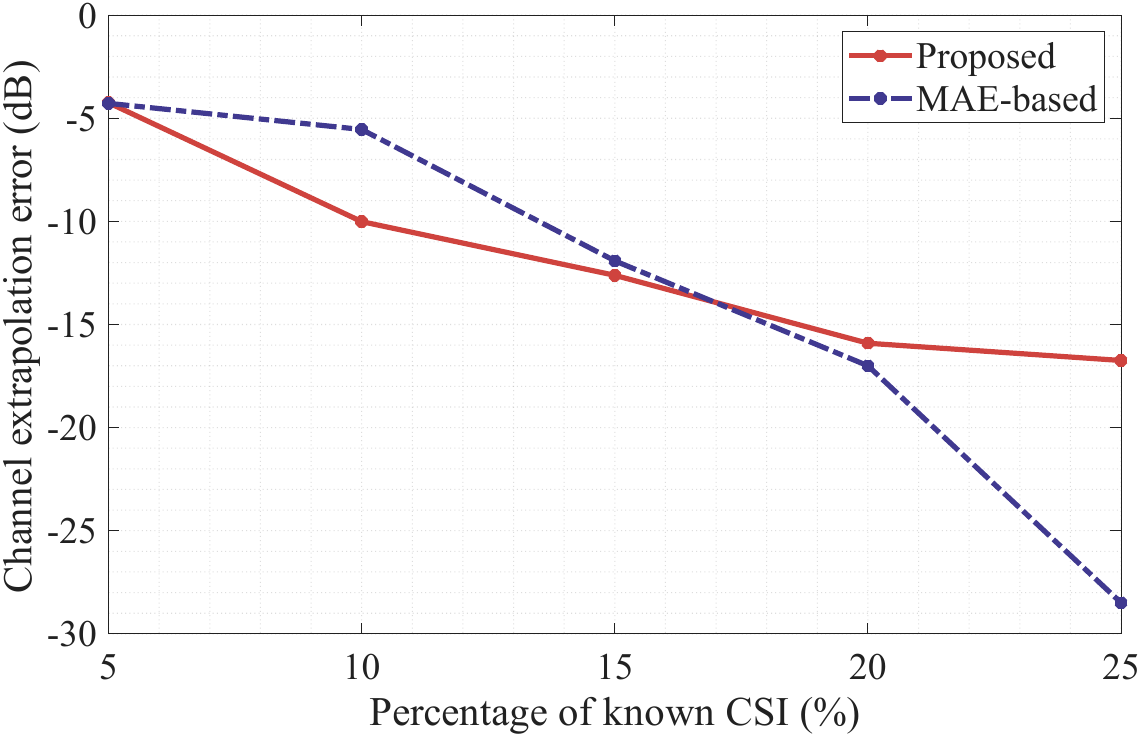}
\caption{The testing performance of the proposed model and MAE-based baseline for channel extrapolation.}
\label{fig:training_testing_CE}
\end{figure}
\begin{figure}[!t]
\centering\includegraphics[width=0.5\textwidth]{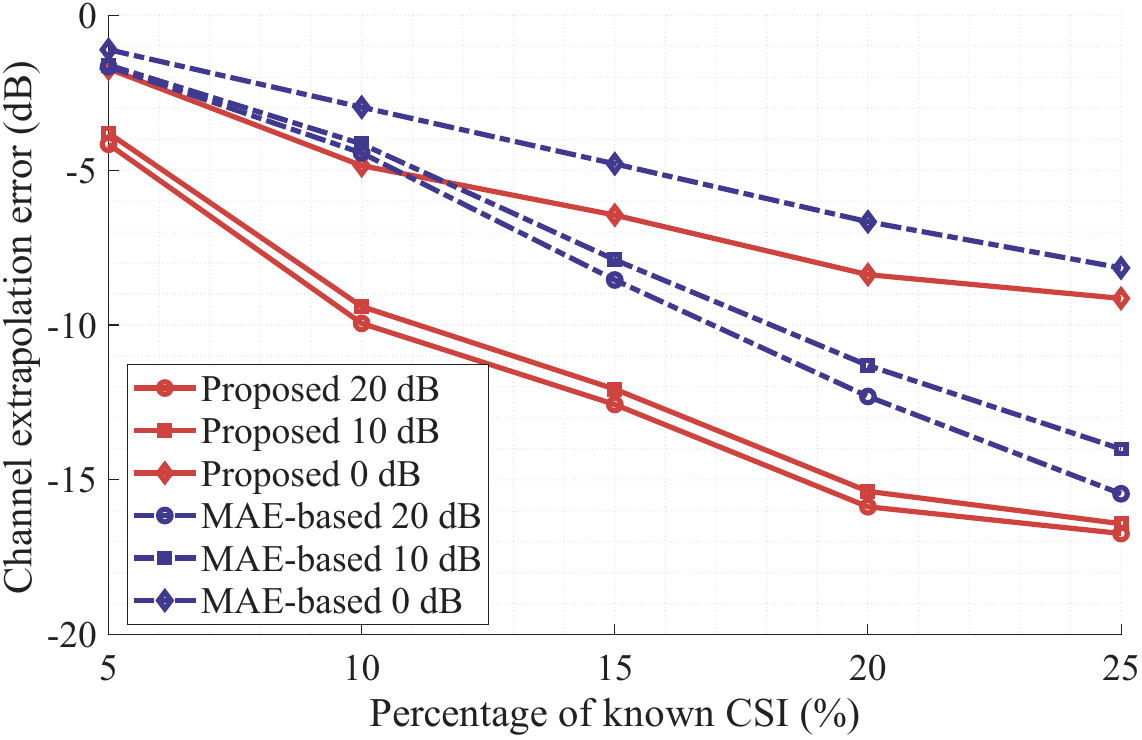}
\caption{The robustness the proposed model and MAE-based baseline at various SNR levels for channel extrapolation.}
\label{fig:testing_CE}
\end{figure}
Fig.~\ref{fig:training_testing_CE} illustrates the core performance trend under nominal high-SNR conditions. The proposed JEPA-CFM consistently achieves lower extrapolation error than the conventional MAE baseline, with the largest gains appearing in the most practically relevant regime of highly sparse observations (5\%--15\% known CSI). This advantage originates directly from the JEPA latent-prediction objective, which guides the encoder toward abstract, high-level spatial structures instead of attempting noisy pixel-level coefficient reconstruction. At moderate observation densities (15\%--20\%), the two approaches reach comparable accuracy. At 25\% known CSI the pure MAE baseline shows a steeper error reduction, as expected when abundant visible data favors direct reconstruction. These results underscore the foundation-model design philosophy: JEPA-CFM is intentionally optimized for robust, data-efficient generalization under realistic FAS constraints rather than for peak reconstruction accuracy when observations are plentiful.

Fig.~\ref{fig:testing_CE} extends the analysis by examining robustness across three representative SNR levels (20 dB, 10 dB, and 0 dB), thereby simulating the thermal noise, interference, and deep fading conditions typical of mobile 6G deployments. Across all tested SNRs, JEPA-CFM maintains its performance lead in the low-observation regime that dominates real-world operation. Moreover, the gap between the JEPA-CFM and MAE curves remains stable or widens slightly as SNR decreases, confirming that the combination of latent predictive learning and sliced isotropic Gaussian regularization effectively mitigates noise amplification that affects conventional reconstruction-based methods. Both models exhibit the expected monotonic improvement with increasing known-CSI percentage and higher SNR, yet the JEPA-CFM curves are consistently lower and flatter at low observation ratios, demonstrating superior noise resilience and better exploitation of spatial correlations even under 0 dB conditions.
\subsubsection{Wireless positioning}
\begin{figure}[!t]
\centering\includegraphics[width=0.5\textwidth]{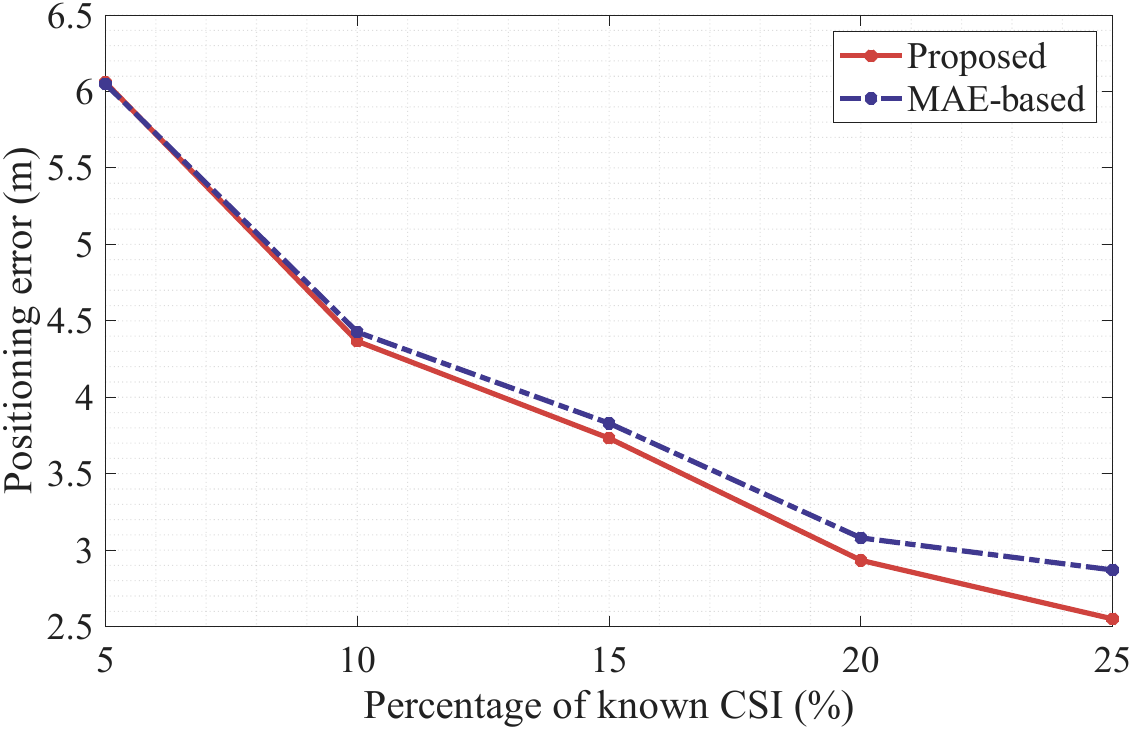}
\caption{The testing performance of the proposed model and MAE-based baseline for positioning.}
\label{fig:training_testing_p}
\end{figure}
\begin{figure}[!t]
\centering\includegraphics[width=0.5\textwidth]{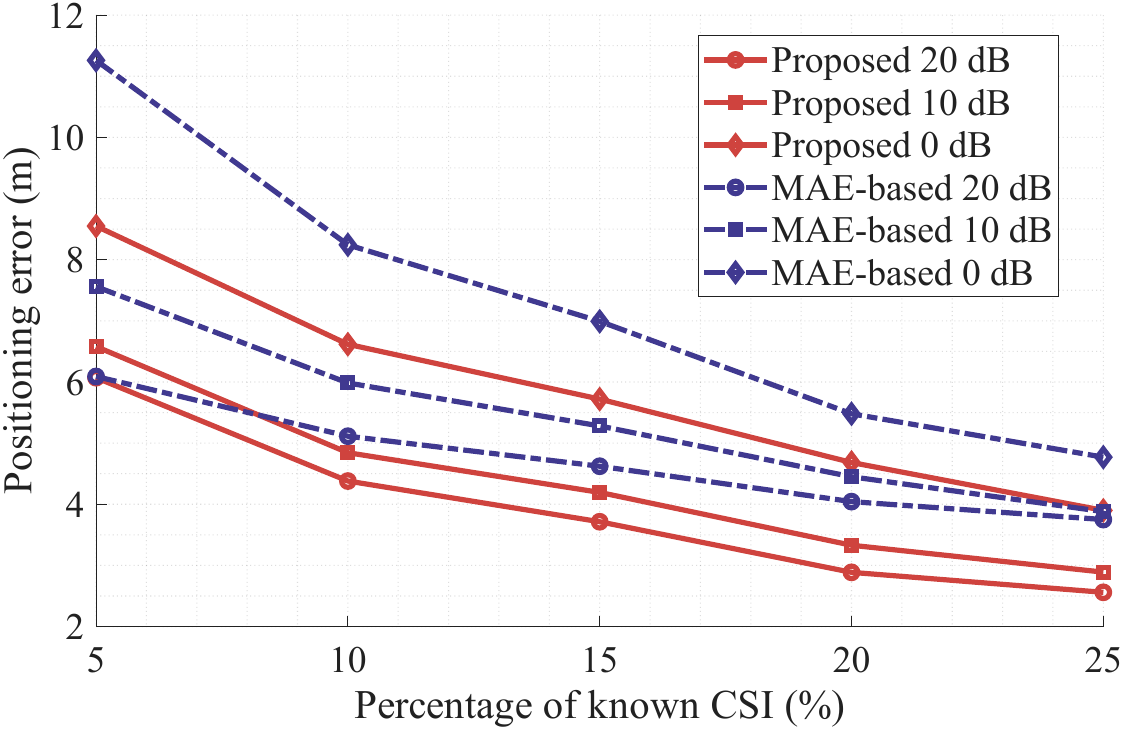}
\caption{The robustness the proposed model and MAE-based baseline at various SNR levels for positioning.}
\label{fig:robustness_testing_CE}
\end{figure}
Fig.~\ref{fig:training_testing_p} illustrates the mean positioning error under nominal (noise-free) conditions as a function of the percentage of known CSI. As shown, the proposed JEPA-CFM model and the MAE-based baseline exhibit very similar performance across the entire range of observation densities. The two curves are nearly overlapping, particularly from 10\% to 25\% known CSI, indicating no significant difference in localization accuracy between the two methods in the absence of noise. Even in the low-observation regime (5\%--15\% known CSI), the gap remains minimal. At 25\% known CSI, both approaches achieve positioning errors around 2.7--3.0 m, with the proposed model showing only a marginal advantage. This suggests that under ideal channel conditions, the predictive latent objective and spatial embeddings in JEPA-CFM do not provide a substantial edge over the reconstruction-based MAE baseline for this positioning task.

Fig.~\ref{fig:robustness_testing_CE} further examines the robustness of both models under varying SNR levels (20 dB, 10 dB, and 0 dB). Here, the performance gap becomes more apparent. At each SNR, the JEPA-CFM curves consistently lie below those of the MAE-based baseline, with the advantage being most visible in the low-observation regime (5\%--15\% known CSI). As SNR decreases to 0 dB, the MAE baseline suffers more pronounced degradation, while the proposed model maintains relatively better performance. Both methods show the expected monotonic decrease in positioning error with increasing known-CSI percentage and higher SNR. However, the JEPA-CFM curves are generally flatter, highlighting improved noise resilience stemming from its latent predictive training and FAS-specific spatial embeddings, which appear to better suppress noise amplification compared to direct reconstruction.

\section{Conclusions}
\label{sec:conclusions}
This paper proposed JEPA-CFM, the first channel foundation model for FAS built upon the JEPA. Unlike conventional masked autoencoders that reconstruct raw complex channel coefficients at the pixel level, JEPA-CFM learns robust, high-level latent representations by predicting abstract embeddings of unobserved channel segments in a compact feature space. To further stabilize training and prevent representation collapse under the strong spatial correlations inherent in compact FAS apertures, the pre-training objective integrates three carefully balanced losses: the standard masked autoencoder reconstruction loss, the proposed JEPA latent prediction loss, and a SIGReg term. This multi-objective design enables a unified, self-supervised encoder that generalizes effectively across downstream tasks with minimal adaptation overhead. Extensive ray-tracing simulations in the realistic DeepMIMO urban scenario confirm the practical advantages of the proposed framework. For channel extrapolation, JEPA-CFM consistently achieves lower normalized mean squared error than the pure masked autoencoder baseline, with the performance gap being most pronounced under severe pilot sparsity (5 \%–15 \% known CSI) and across a wide range of SNR conditions (0 dB, 10 dB, and 20 dB). The latent predictive objective and SIGReg regularization together provide superior noise resilience and better exploitation of spatial correlations compared with pixel-level reconstruction methods. For the wireless positioning task, the pre-trained encoder with a lightweight regression head delivers sub-3 m mean localization error at only 25 \% known CSI, demonstrating strong geometric awareness even when most ports remain unobserved.

\ifCLASSOPTIONcaptionsoff
  \newpage
\fi
 \small
\bibliographystyle{IEEEtran}
\bibliography{reference.bib}
\end{document}